\documentstyle[psfig,mnras_cite,times]{mn}
\newcommand{\nn}{\nonumber \\}
\newcommand{\ls}{\mathrel{\raise1.16pt\hbox{$<$}\kern-7.0pt %  <
\lower3.06pt\hbox{{$\scriptstyle \sim$}}}}         %  ~
\newcommand{\gs}{\mathrel{\raise1.16pt\hbox{$>$}\kern-7.0pt %  >
\lower3.06pt\hbox{{$\scriptstyle \sim$}}}}         %  ~

\long\def\comment#1{}

\def\fun#1#2{\lower3.6pt\vbox{\baselineskip0pt\lineskip.9pt
  \ialign{$\mathsurround=0pt#1\hfil##\hfil$\crcr#2\crcr\sim\crcr}}}

\def\gap{\mathrel{\mathpalette\fun >}}
\def\ba{\begin{eqnarray}}
\def\ea{\end{eqnarray}}
\def\be{\begin{equation}}
\def\ee{\end{equation}}
\def\nn{\nonumber \\}
\def\vk{{\bf k}}

\def\eps{\epsilon}

\begin{document}
\title[Tests for primordial non-Gaussianity]{Tests for
primordial non-Gaussianity}
\author[Verde, Jimenez, Kamionkowski \& Matarrese]{Licia Verde$^{1}$, Raul
Jimenez$^2$, Marc Kamionkowski$^3$ $\&$ Sabino Matarrese$^{4}$ \\
$^1$ Department of Astrophysical Sciences, Princeton University, Princeton,
NJ 08540--1001 USA (lverde@astro.princeton.edu) \\
$^2$ Department of Physics and Astronomy, Rutgers University, 136
Frelinghuysen Road, Piscataway, NJ 08854--8019 USA (rauj@physics.rutgers.edu)\\
$^3$ Mail Code 130--33, California Institute of Technology,
Pasadena, CA 91125 USA (kamion@tapir.caltech.edu)\\
$^4$ Dipartimento di Fisica
"Galileo Galilei", via Marzolo 8, I-35131 Padova,
Italy. (matarrese@pd.infn.it)}

\maketitle

\begin{abstract}
We investigate the relative sensitivities of several tests for
deviations from Gaussianity in the primordial distribution of
density perturbations.  We consider models for non-Gaussianity
that mimic that which comes from inflation as well as that which 
comes from topological defects.  The tests we consider involve
the cosmic microwave background (CMB), large-scale structure
(LSS), high-redshift galaxies, and the abundances and properties 
of clusters.  We find that the CMB is superior at finding
non-Gaussianity in the primordial gravitational potential (as
inflation would produce), while observations of high-redshift
galaxies are much better suited to find non-Gaussianity that
resembles that expected from topological defects. 
We derive a
simple expression that relates the abundance of high-redshift
objects in non-Gaussian models to the primordial skewness.
\end{abstract}


\section{Introduction}
%%%%%%%%%%%%%%%%%%%%%%%

Now that cosmic-microwave-background (CMB) experiments
\cite{boomerang3,maxima,balbi,lange} have verified the inflationary predictions of a flat
Universe and structure formation from primordial adiabatic perturbations, we
are compelled to test further the predictions of the simplest
single-scalar-field slow-roll inflation models and to look for possible
deviations.  Measurements of the distribution of primordial density
perturbation afford such tests.  If the primordial perturbations are due
entirely to quantum fluctuations in the scalar field responsible for inflation
(the ``inflaton''), then their distribution should be {\em very} close to 
Gaussian 
(e.g., \pcite{Guthpi82,Starobinski82,BST83,FRS93,GLMM94,Gangui94,WK00,Ganguimartin99}).
However, multiple-scalar-field models of inflation allow for the possibility
that a small fraction of primordial perturbations are produced by quantum
fluctuations in a second scalar field. If so, the distribution of these
perturbations could be non-Gaussian (e.g.,
\pcite{allengrinsteinwise87,kofpog88,SBB89,LM97,Pee99a,Pee99b,Salopek99}).
Moreover, it is still possible that some component of primordial perturbations
are due to topological defects or some other exotic causal mechanism
\cite{bouchet}, and if so, their distribution should be non-Gaussian (e.g.,
\pcite{vilke85,vach86,hillscrammfry89,Tur89,albsteb92}).  Detection of any
non-Gaussianity would thus be invaluable for appreciating the nature of the
ultra-high-energy physics that gave rise to primordial perturbations.  Ruling
such exotic possibilities in or out will also be necessary to test the
assumptions that underly the new era of precision cosmology.

There are several observables that can be used to look for
primordial non-Gaussianity.  CMB maps probe cosmological
fluctuations when they were closest to
their primordial form, and many authors have developed various mathematical
tools to test the Gaussian hypothesis.
The statistics of present-day large scale structure (LSS) in the
Universe can also be used
(e.g., \pcite{CMLMM93,LS93,Lokas95,Chodbouch96,SP96,Durreretal00,VH00}).
The properties and abundances of the most massive and/or
highest-redshift objects in the Universe also contain precious
information about the nature of the initial conditions
(e.g., \pcite{COS98}; Robinson, Gawiser \& Silk 1999; Robinson, Gawiser \&
Silk 2000; \pcite{Willick00,MVJ00} (MVJ); \pcite{VKMB}).  
In Verde et al. (2000a; VWHK00), the relative sensitivities of the CMB and LSS
to several broad classes of primordial non-Gaussianity were compared, and it
was found that forthcoming CMB maps can provide more sensitive probes of
primordial non-Gaussianity than galaxy surveys.  Here we extend the results of
that paper to include comparisons to the abundances of high-redshift galaxies
as well as the abundance and properties of clusters.  One of our original aims
was to determine whether any of these probes would be able to detect the
miniscule deviations from Gaussianity that arise from quantum fluctuations
in the inflaton; unfortunately, we have been unable to find any.
Nevertheless, some detectable deviations from Gaussianity are conceivable with
multiple-field models of inflation and/or some secondary contribution to primordial
perturbations from topological defects. We will follow VWHK00 and parameterize
the primordial non-Gaussianity with a parameter that can be dialed from zero
(corresponding to the Gaussian case) for two different classes of
non-Gaussianity.  We will then compare the smallest value for the parameter
that can be detected with each of the different approaches.
%\vspace*{-0.3cm}
\section{The method}
\subsection{Models for primordial non-Gaussianity}
%NEW
There are infinite types of possible deviations from Gaussianity, and it is
unthinkable to address them all.
 However, we can consider plausible physical
mechanisms that produce small deviations from the Gaussian behavior and thus  
analyze the following two models for the primordial non-Gaussianity
(e.g., Coles \& Barrow 1987, VWHK00, MVJ).  
%END NEW
In the first model, we suppose that the fractional density
perturbation $\delta(\vec x)$ is a non-Gaussian random field
that can be written in terms of a Gaussian random field
$\phi(\vec x)$ through (Model A)
\begin{equation}
     \delta=\phi+\eps_{\rm A}(\phi^2-\langle\phi^2\rangle).
\label{firstngmap}
\end{equation}
In the second model, we suppose that the primordial
gravitational potential $\Phi(\vec x)$ is a non-Gaussian random
field that can be written in terms of a Gaussian random field
$\phi(\vec x)$ through (Model B)
\begin{equation}
     \Phi=\phi+\eps_{\rm B}(\phi^2-\langle\phi^2\rangle).
\label{secondngmap}
\end{equation}
Non-Gaussianity in the density field is then obtained from that
in the potential through the Poisson equation.
Here, $\Phi$ and $\delta$ refer to the {\it primordial}
gravitational potential and density perturbation, before the
action of the transfer function that takes place near
matter-radiation equality.

Although not fully general, these models may be considered as
the lowest-order terms in Taylor expansions of more general
fields, and are thus quite general for small deviations from
Gaussianity.  The scale-dependence of the non-Gaussianity in the 
two models differs.  Model A produces deviations from
Gaussianity that are roughly scale-independent on large scales, while Model B
produces deviations from
non-Gaussianity that become larger at larger distance scales.  
Although we choose these models
essentially in an {\it ad hoc} way, the non-Gaussianity of
Model B is precisely that arising in standard slow-roll 
inflation and in non-standard (e.g., multifield) inflation
(\pcite{Luo94,FRS93,GLMM94,Fanbardeen92}; see also below).  Model
A more closely resembles the non-Gaussianity that would be
expected from topological defects (e.g., VWHK00).
In either case, the lowest-order deviations from non-Gaussianity 
(and those expected generically to be the most easily observed)
are the three-point correlation function (including the
skewness, its zero-lag value) or equivalently the bispectrum, its
Fourier-space counterpart.  It is straightforward to calculate
these quantities for both Models A and B.

\subsection{Cosmic Microwave Background  and Large Scale Structure}

Temperature fluctuations in the CMB come from density
perturbations at the surface of last scatter, so the distribution
of temperature fluctuations reflects that in the primordial
density field.  It is thus straightforward to relate the
density-field bispectra of Models A and B to the bispectrum of
the CMB.  Density perturbations in the Universe today grew 
via gravitational infall from primordial perturbations in the
early Universe, and this process alters the mass distribution 
in a calculable way.  Cosmological perturbation theory allows
the bispectrum for the mass distribution in the Universe today
to be related to that for the primordial distribution.

VWHK00 calculated the smallest values of $\epsilon_{\rm A}$
and $\epsilon_{\rm B}$ that would be accessible with the CMB and
with LSS.  For the CMB calculation, it was assumed that a
temperature map could be measured to the cosmic-variance limit
only for multipole moments $\ell\ls 100$; it was assumed
(quite conservatively) that no information would be obtained
from larger multipole moments.  The LSS calculation were made under the
very optimistic assumption that the distribution of mass could
be determined precisely from the galaxy distribution (i.e., that 
there was no biasing) in a survey of the size of SDSS and/or
2dF.  VWHK00 found that the smallest values of $\epsilon$ that
can be detected with the CMB under these assumptions is
$\epsilon_{\rm A} \sim 10^{-2}$ and $\epsilon_{\rm B} \sim 20$
(\pcite{Komatsuspergel00} including noise and foreground but neglecting dust
contamination found that $\epsilon_{\rm B} \gap 5$ from the Planck experiment
), while the
smallest values measurable with LSS are $\epsilon_{\rm A}\sim 10^{-2}$
and $\epsilon_{\rm B}\sim 10^3$.  More realistically, the galaxy
distribution will be biased relative to the mass distribution,
and this will degrade the sensitivities to nonzero $\epsilon_{\rm A}$
and $\epsilon_{\rm B}$ obtainable with LSS.  VWHK00 thus concluded
that the CMB will provide a keener probe of primordial
non-Gaussianity for the class of models considered.

\begin{figure}
\centerline{
\psfig{figure=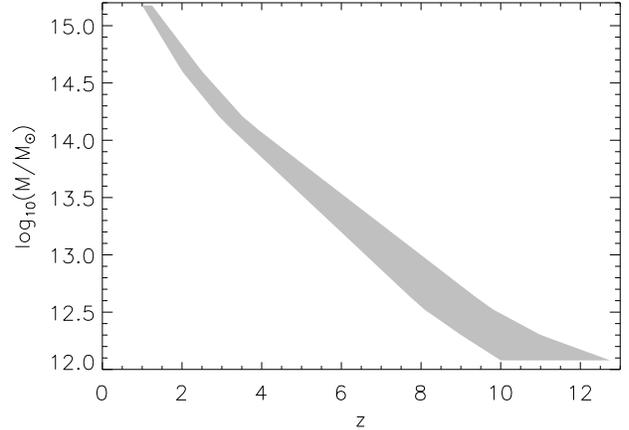,height=6.5cm,angle=0}}
\caption{$M_{\rm max}$ as a function of redshift. At a given redshift one
should only consider those masses ($\le M_{\rm max}$) for which at least one
object is expected in the whole sky for Gaussian initial conditions. The shaded region encloses predictions for $M_{\rm max}(z)$ from
different mass functions in the literature; we adopted the currently favored
cosmological model with parameters: $\Omega_0=0.3$, $\Lambda_0=0.7$, $h=0.65$,
$\sigma_8=0.99$ and transfer function of Sugiyama (1995) with
$\Omega_b=0.015/h^2$ ($\Lambda$CDM).\label{figure1}}
\end{figure}

\subsection{High-redshift and/or massive objects}

According to the Press-Schechter theory, the abundance of high-redshift and/or
massive objects is determined by the form of the high-density tail of the
primordial density distribution function.  A probability distribution function
(PDF) that produces a larger number of $>3\sigma$ peaks than a Gaussian
distribution will lead to a larger abundance of rare high-redshift and/or
massive objects.  
%*NEW*
Since small deviations from Gaussianity have deep impact on
those statistics that probe the tail of the distribution (e.g. \pcite{Fry86}, MVJ), rare high-redshift and/or
massive objects should be powerful probes of primordial non Gaussianity.
 The number densities of high-redshift galaxies and/or of
clusters (at either low or high redshifts) provides a very sensitive probe
of the PDF.  Since the Gaussian tail is decaying exponentially at higher
densities, even a small deviation from Gaussianity can lead to huge
enhancements in the number densities.

The non-Gaussianity parameters $\epsilon_{\rm A, B}$ are effectively ``tail
enhancement'' parameters (c.f., MVJ)\footnote{In fact, when looking on a
particular scale, it is always possible to parameterize the deviation of the
PDF  from Gaussianity, with some ``effective'' $\epsilon_{\rm A}$ or
$\epsilon_{\rm B}$, if the PDF is not too non-Gaussian. It is easy to
understand this statement if one thinks in terms of skewness. Physical
mechanisms that produce non-Gaussianity generically produce non-zero skewness in the PDF for
the simple reason that underdense region cannot be more empty than voids
while overdense regions can become arbitrarily overdense. Skewness can be
scale dependent, but for a given value of the skewness there is one-to-one
correspondence to $\epsilon_{\rm A,B}$ parameters (see the Appendix).}  
%*END NEW

In order to determine the minimum value of $\eps_{\rm A,B}$ that can be
detected using high-redshift objects, one needs to compute by how much the
observed number density of objects changes with respect to the Gaussian
case when the primordial field is described by equations
(\ref{firstngmap}) and (\ref{secondngmap}).
We calculate this enhancement using the results for the mass function for
mildly non-Gaussian initial conditions obtained analytically in
MVJ.  
Conservatively, we make the assumption that objects form at
the same redshift at which they are observed ($z_c=z$); since for some objects the
dark halo will have collapsed before we observe them, the assumption gives a
{\em lower limit} to the amount of non-Gaussianity.

The directly observed quantity, however, is not the mass
function, but is $N(\geq M,z)$, the %NEW
total number of objects --in the survey area-- of mass $\geq M$ that collapse at
redshift $z$.  In fact it is extremely difficult to obtain an accurate
estimate of the mass of high-redshift objects, what is a more robust quantity
is the minimum mass that these objects must have in order to be detected at
that redshift. 
%END NEW
This quantity is related to the mass function, $n(M,z)$, by 
\be N(\geq M, z)=\int_M^{\infty}n(M,z)dM\;. 
\ee

In calculating the enhancement of high-redshift objects due to primordial
non-Gaussianity, we restrict ourselves to consider, at any given redshift,
only those masses, $M\le M_{\rm max}(z)$ for which at least one object is expected
in the whole sky for Gaussian initial conditions 
%NEW
($N(\ge M_{\rm
max},z)=1$ in $4\pi$ radians)\footnote{This choice for the threshold $N(\ge M_{\rm
max},z)=1$ is motivated by the following considerations. Of course it is not robust to detect a non-Gaussianity that suppresses the
number of objects with respect to the Gaussian prediction, since one can
always argue that one did not look hard enough, or that the objects are there
but are  somewhat ``invisible''.
So we set to detect a non-Gaussianity that enhances the number of objects
relative to the Gaussian case. If within Gaussian initial conditions
we expect $N(>M,z)\sim 0$ in the whole sky, and observations find $N(>M,z)>1$
in the survey area, we can say that we have detected non-Gaussianity.
However, the non-Gaussianity (or tail enhancement)  parameter  is directly related to the
ratio of observed $N_{ng}(>M,z)$ to the Gaussian predicted $N(>M,z)$
(see Eq. 4). Obviously this ratio is well defined for any $N_{\rm ng}>0$ and $N>0$,
but the observed  $N_{\rm ng}$ can only be an integer $\ge 1$. The tail enhancement parameter
will then make $N_{\rm ng} \ge N$ (and we consider only cases where $N_{\rm
ng} \gap 10 N$). It is reasonable therefore to consider only
those masses and redshift for which the theoretical prediction for the
Gaussian $N$ is $\ge 1$.}. 
%END NEW
This is illustrated in Fig. 1 for a $\Lambda$CDM model (hereafter
we adopt the currently favored cosmological model with parameters:
$\Omega_0=0.3$, $\Lambda_0=0.7$, $h=0.65$, $\sigma_8=0.99$ and transfer
function of \scite{Sugiyama95} with $\Omega_b=0.015/h^2$) where the shaded
region encloses predictions for $M_{\rm max}(z)$ from different mass functions
(e.g., \pcite{PS74,LeeShandarin98,ShethTormen99,Jenkinsetal00}).

Given the rapidly dying tail of the Gaussian PDF, small
uncertainties in the mass determination of high-redshift objects
could lead to overestimate the value of $\eps_{\rm A,B}$.  An
overabundance of galaxies of estimated mass $M_e$, which 
in principle can be attributed to a non-zero value of
$\epsilon_{\rm A,B}$, can also be explained under the hypothesis 
of Gaussian initial conditions if the actual galaxy mass
$M_{\rm true}$ is $M_{\rm true}< M_e$.  We thus include
conservative values for the uncertainty $\Delta M$ in the mass
determination of high-redshift objects and we then
calculate the minimum change $\Delta N$ in the number density of objects
over the Gaussian case that cannot be attributed to the
uncertainty in the mass determination.
For a given uncertainty in the mass, this can be computed by
using the standard Press-Schechter (PS) theory \cite{PS74}.
Observationally it is difficult to measure the mass of
high-redshift clusters with accuracy better than 30\%, with
either weak lensing or the X-ray temperature, and of
high-redshift galaxies better than a factor 2 ($\Delta M=M$;
at least of their stellar mass).  
Although the calculations in this section are obtained using 
the standard PS theory, our
conclusions will be essentially unchanged if we had used
modified PS theories
(e.g., \pcite{LeeShandarin98,ShethTormen99,ShethMoTormen99,Jenkinsetal00}; see
below).

With the mass uncertainties discussed above, we obtain that the
minimum $\Delta N$ that cannot be attributed to $\Delta M$ is a
factor 10 for clusters and a factor 100 for galaxies (see, e.g., Fig. 6 of
MVJ00).

We therefore estimate the minimum $\eps_{\rm A,B}$ that can be
measured from the abundance of high-redshift objects as the one that
corresponds respectively to a factor 100 and 10 change in the
observed number density of objects ($N(\geq M,z)$) over the
Gaussian case.
This condition can be written as
\be
N_{ng}(\geq M,z)/N(\geq M, z)\equiv R(M,z)\geq R_* ,
\ee 
where $N$ is obtained using the Gaussian mass function while $N_{ng}$ is
obtained using the non-Gaussian mass function as in MVJ, and $R_*$
is set to be 100 for galaxies and 10 for clusters.

For small primordial non-Gaussianity (i.e., for small values for $\eps_{\rm
A,B}$), it is possible to derive an expression for $R(M,z)$ using the
analytical approximation for the mass function $n_{ng}$ found in MVJ. Doing so
we find

\begin{equation}
R(M,z)\simeq\frac{\int_M^{\infty}(\sigma_MM)^{-1}
\exp[-\frac{\delta_*^2(z_c)}{2\sigma_M^2}]F(M,z_c,\eps_{\rm A,B})dM}
{\int_M^{\infty}(\sigma_MM)^{-1}\exp[-\frac{\delta_c^2(z_c)}
{2\sigma_m^2}]\left|\frac{d\sigma_M}{dM}\right|dM}\;.
\label{ratio}
\end{equation}
Here,
\ba
F(M,z_c,\eps_{\rm A,B})& =
&\left|\frac{\delta_c(z_c)}{6\sqrt{1-S_{3,M}\delta_c(z_c)/3}} \frac{d
S_{3,M}}{dM} \right. \nn
 & &\left.
+\frac{\sqrt{1-S_{3,M}\delta_c(z_c)/3}}{\sigma_M}\frac{d\sigma_M}{dM}\right|,
\ea
and
\be
\delta_*(z_c)=\delta_c(z_c)\sqrt{1-S_{3,M}\delta_c(z_c)/3},
\ee
\be
\delta_c(z_c)=\Delta_c/D(z_c), 
\ee
where $D(z_c)$ is the linear-theory growth factor, and $\Delta_c$ is the
linear extrapolation of the over-density for spherical collapse. 

In the formulas above, $S_{3,M}$ denotes the primordial skewness,
\be
S_{3,M}=\eps_{\rm A,B}\mu^{(1)}_{3,M}/\sigma^{2}_M,
\ee
where the expressions for $\mu_{3,M}^{(1)}$ and $\sigma^2_M$ can be found
in MVJ section 3.2 equations (37) and (38).
\begin{figure*}
\begin{center}
\setlength{\unitlength}{1mm}
\begin{picture}(90,55)
\includegraphics{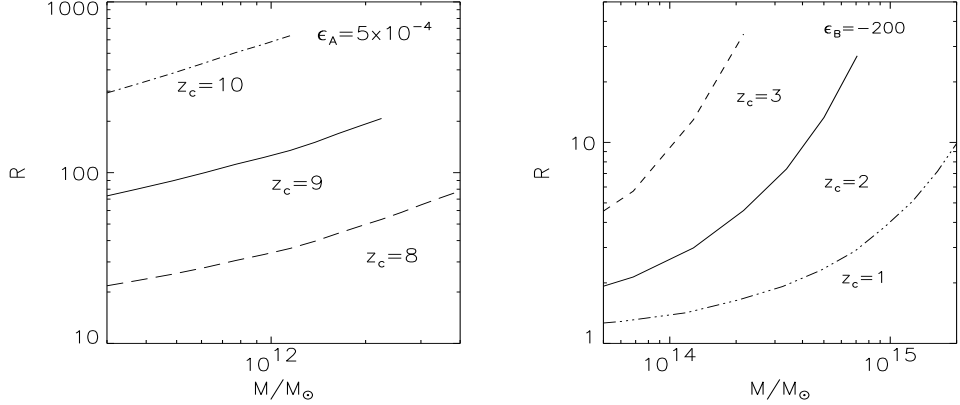}
\end{picture}
\end{center}
\caption{\label{figure2} Ratio $R(M,z)=N_{ng}(\geq M,z)/N(\geq M,z)$ for
galaxies at redshift $z=8, 9$ and 10 for $\eps_{\rm A}=5\times 10^{-4}$ (left panel)
and clusters at redshift $z=1 ,2$ and 3 (right panel), for $\eps_{\rm B}=200$ as a
function of $M$. Lines are plotted only for masses where, for Gaussian initial
conditions, one would expect to observe at least one object in the whole
sky with the most conservative estimate
(see Fig.~\ref{figure1}). Note that these
high-redshift objects represent 3- to 5-$\sigma$ peaks. The values for the
number density enhancement $R$ that
can safely be attributed to primordial non-Gaussianity are $R=100$ for galaxies
(left panel) and $R=10$ for clusters (right panel). See text for details.}  
\end{figure*}
However, for $S_3\gap 1/\delta_c(z_c)$, the mass function $n_{ng}(M,z)$ has to
be evaluated numerically and equation (\ref{ratio}) is not valid.

For the cosmological model considered here and the redshifts of interest,
the quantity $\Delta_c$ takes a nearly constant value ($\approx 1.686$) in the PS theory. 
A better fit to the mass function of halos in 
high-resolution N-body simulations is however obtained by lowering $\Delta_c$
for rare objects and giving it 
an extra mass and redshift dependence (\pcite{ShethTormen99}; \pcite{Bodeetal00}), as motivated by ellipsoidal
collapse (e.g., \pcite{LeeShandarin98,ShethMoTormen99}).

It is possible to understand the effect of a lower $\Delta_c$ by the
following argument. For rare fluctuations such as high-redshift objects one
is probing the mass function above the knee. Since the mass function drops
very rapidly as $M$ increases we can approximate $N(>M,z_c)\sim n(M,z_c)M$.
It is then possible to obtain an analytic expression for
$r(M,z_c)\equiv n_{ng}(M,z_c)/n(M,z_c)\sim R(M,z_c)$ if the primordial non-Gaussianity is small:
\be
r(M,z_c) \!\simeq \!\exp{\!\left[\frac{\Delta_c^3 S_3}{6 \sigma_{\rm M}^2}\right ]}
 \left|
\frac{\delta_c}{6 \sqrt{1-\frac{S_3 \delta_c}{3}}} \frac{d S_3}{d
\sigma_M} +
 \sqrt{1-\frac{S_3 \delta_c}{3}} \right|.
\ee  

For a given mass $M$, $r(M,z_c)$ slowly decreases when lowering $\Delta_c$,
slightly damping the effect of non-Gaussianity. For example when lowering
$\Delta_c$ from the value we assume here 1.686, to the value $\approx 1.5$---appropriate to
fit  the numerical mass function of 
\scite{ShethTormen99} for the range 
of masses and redshifts considered here---$r(M,z_c)$ decreases by less 
than a factor 2. However this effect is compensated by the fact that, by lowering
$\Delta_c$, objects are created more easily also with Gaussian initial
conditions, and it is therefore possible to consider objects of higher $M$
and/or $z$, where the effect of non-Gaussianity is bigger.  In summary, the
conclusions obtained by assuming $\Delta_c=1.686$ will not be substantially
modified.  

It is important to note that for Model A, the primordial skewness has the same
sign as $\eps_{\rm A}$, while for Model B the primordial skewness has the opposite sign
of that of $\eps_{\rm B}$.  In detecting non-zero $\eps_{\rm A,B}$ from CMB maps, the sign
of the skewness does not influence the accuracy of the detection of
non-Gaussianity, but, when using the abundance of high-redshift objects  the
sign of the skewness matters. 
Only a positively skewed primordial distribution will generate more
high-redshift objects than predicted in the Gaussian
case. Although a negatively
skewed probability distribution will generate fewer objects than the Gaussian
case, a decrement might be difficult to attribute exclusively to
a negatively skewed distribution. Therefore in the
following we will consider only negative $\eps_{\rm B}$ and positive $\eps_{\rm A}$.

\subsubsection{Cluster size-temperature distribution}
\scite{VKMB} showed
that the size-temperature (ST) distribution of clusters is fairly sensitive to the
degree of primordial non-Gaussianity. If clusters are created from rare
Gaussian peaks, the spread in formation redshift should be small and so should
the
scatter in the ST distribution. Conversely, if the probability
distribution function has long non-Gaussian tails, then clusters of a given mass we
observe today should have a broader formation redshift distribution and thus a
broader ST relation.  In \scite{VKMB}, the non-Gaussianity
considered is a log-normal distribution; it is not strictly equivalent to
Models A or B considered here. However, for small deviations from Gaussianity,
the two models can be identified if, for a given scale, they produce the same
skewness in the density fluctuation field.  We thus find that in the
$\Lambda$CDM model the minimum $\epsilon_{\rm A}$ and $\epsilon_{\rm B}$ detectable with
the ST distribution method are $3\times 10^{-3}$ and $500$
respectively. These estimates assume that the cosmology and $\sigma_8$ are
well known, but use only the local cluster data set of \scite{M00}. Of course, with improved observational data, the ST method
could probably yield stronger constraints.

\section{Results}

We find that the non-Gaussianity of Model A has a bigger effect on
high-redshift galaxies than on high-redshift clusters. This can be understood
for the following reason. For Model A the skewness $S_{3,M}$ is approximately
scale independent ($dS_{3,M}/dM=0$). Thus, as found in MVJ, the mass function
for non-Gaussian initial conditions is obtained from the PS mass function for
Gaussian initial conditions replacing $\delta_c(z_c)
\longrightarrow \delta_*(z_c)$. The effect of a non-zero skewness is therefore
to lower the effective threshold for collapse  thus allowing more objects to
be created. For a given $S_3$, $\delta_*(z_c)$ is a monotonically decreasing
function of $z_c$. Since galaxies can be observed at $z_c$ much bigger than
that of clusters, the effect is bigger. On the other hand, clusters are better
probes than galaxies for Model B. In fact, for Model B the induced skewness in
the density field is scale dependent and the effect of
non-Gaussianity is roughly the same for galaxies with $ 8<z<10$ and clusters
with $1<z<3$. However since mass determinations are more accurate for clusters
than for galaxies, we have $R_{*,{\rm clusters}}<R_{*,{\rm
galaxies}}$: clusters are therefore better probes.

In Fig.~\ref{figure2} we show the ratio $R=N_{ng}(\geq M,z)/N(\geq M, z)$
(cf. eq. (5)) for
galaxies at redshift $z=8, 9$ and 10 for $\eps_{\rm A}=5\times 10^{-4}$ (model A, left panel) and
clusters at redshift $z=1 ,2$ and 3  for $\eps_{\rm B}=-200$ (model B, right panel),
as a function of $M$.  Lines are plotted only for masses where, for Gaussian initial
conditions, one would expect to observe at least one object in the whole sky
with the most conservative estimate (see Fig.~\ref{figure1}).  Note that those
high-redshift objects represent 3- to 5-$\sigma$ peaks. If we now require $R(M,z_c) > R_*$, we deduce that
the minimum detectable deviation from Gaussian initial conditions will be
$\eps_{\rm A}\sim 5 \times 10^{-4}$ (from high-redshift galaxies) and
$|\eps_{\rm B}|\sim
200$ (from high-redshift clusters). 
We also estimate that an uncertainty of 10\% on $\sigma_8$ would propagate
into an uncertainty of 25\% in $\epsilon_{\rm B}$ (from clusters)  and of
70\% in $\epsilon_{\rm A}$ (from galaxies).  

The minimum $\epsilon_{\rm B}$ detectable from high-redshift cluster abundances is much larger than the value
that can be measured from the CMB ($\eps_{\rm B} \sim 5$ to $20$ for Planck
data), while for
$\eps_{\rm A}$, high-redshift galaxies are much better probes than the CMB,
which can only detect $\eps_{\rm A} \sim 10^{-2}$.

We therefore conclude that {\em if} future NGST or 30- to 100-m ground-based
telescope observations of high-redshift galaxies yield a significant number of
galaxies at $z \sim 10$ {\em and} are able to determine their masses within a
factor 2, these observations will perform better than CMB maps in constraining
primordial non-Gaussianity of the form of Model A with positive
$\eps_{\rm A}$. Conversely, forthcoming CMB maps will constrain deviations from
Gaussianity in the initial conditions much better than observations of
high-redshift objects for Model B (with positive {\em and} negative value for
$\eps_{\rm B}$) and for Model A with negative $\eps_{\rm A}$.

\subsection{Slow-roll parameters and primordial skewness}
The type of non-Gaussianity of Model B is particularly interesting because
initial conditions set from standard inflation show deviations from
Gaussianity of this kind.  In fact, it is possible to relate the two slow roll
parameters,

\begin{equation}
\eps_*=\frac{m_{Pl}^2}{16
\pi}\left(\frac{V'}{V}\right)^2 \;,\;\;\;
\eta_*=\frac{m_{Pl}^2}{8\pi}\left[\frac{V''}{V}-\frac{1}{2}\left(\frac{V'}{V}\right)^2\right],
\label{slowrollparam}
\end{equation}
to the non-Gaussianity parameter $\eps_{\rm B}$. In equation (\ref{slowrollparam})
$m_{Pl}$ is the Planck mass, $V$ denotes the inflaton potential and $V'$ and
$V''$ the first and second derivatives with respect to the scalar field. 
The skewness $S_3$ for $\Phi_{\rm B}$,
$S_{3,\Phi}=\langle\Phi_{\rm B}^3\rangle/\langle\Phi_{\rm B}^2\rangle^2$, can be
evaluated following a similar calculation of \scite{BK99}, obtaining

\be
S_{3,\Phi}=2\eps_{\rm B}\times 3[1+\gamma(n)],
\ee
where $\gamma(n) \ll 1$ and weakly depends on $n$ if $n<0$, but diverges for
$n>0$.  For a scale-invariant matter density power spectrum, $n=-3$,
$\gamma(n)=0$, and so $S_{3,\Phi}=6\eps_{\rm B}$.

We can then compare this expression with the value for the skewness parameter
for the gravitational potential arising from inflation to infer the
magnitude of $\eps_{\rm B}$.
Gangui et al. (1994) calculate the CMB skewness for the Sach-Wolfe effect
$S_2$ in several inflationary models; $S_2$ is related to $S_{3,\Phi}$ by
$S_2=S_{3,\Phi}A_{sw}^{-1}$ where $A_{sw}=1/3$.
From this it follows that $S_2=3S_{3,\Phi}=18\eps_{\rm B}$.
The condition for slow roll from Gangui et al. 1994 is $S_2\leq 20$; 
thus,  $\eps_{\rm B} \leq 1$, and the  relation to the slow-roll parameters is
(cf., \pcite{WK00}) 
\begin{equation}
\eps_{\rm B}=(5/2)\eps_*-(5/3)\eta_*. 
\end{equation}
Since this combination of the slow-roll parameters is different from the
combination that gives the spectral slope $n$ of the primordial power spectrum
($n=2\eps_*-6\eta_*+1$), in principle, if $\eps_{\rm B}$ could be measured with an
error $\ll 1$, it would be possible to determine the shape of the inflaton
potential through eq. (\ref{slowrollparam}).
However, from the present analysis, an error of $\eps_{\rm B}$ of about an order
of magnitude larger seems to be realistically  achievable. 

\begin{table}
\begin{center}
\begin{tabular}{llllllll}
observable & min. $|\epsilon_{\rm A}|$  & min. $|\epsilon_{\rm B}|$  \\
\hline
CMB   &   $10^{-3}\sim 10^{-2}$ &   $ 20$\\
LSS   &   $10^{-2}$  &   $ 10^3\sim 10^4$\\
High z obj. &  $ (+) 5\times 10^{-4}$ (gal.) & (--) 200 (clusters) \\
ST relation  &  $(+) 3\times 10^{-3}$ & (--) 500\\
\end{tabular}
\caption{Minimum $|\epsilon_{\rm A}|$ and $|\epsilon_{\rm B}|$ detectable form different
observables and their sign when  positive skewness is required for detection. For Model A the primordial skewness has the same sign as
$\epsilon_{\rm A}$, while for Model B the primordial skewness has the opposite sign
as $\epsilon_{\rm B}$. In detecting non-zero $\eps_{\rm A,B}$ from CMB maps, the sign
of the skewness does not influence the accuracy of the detection of
non-Gaussianity, but, when using the abundance of high-redshift objects it is
robust to detect non-Gaussianity that produces an excess rather than a defect
in the number density. Only a positively skewed primordial distribution will
generate more high-redshift objects than predicted in the Gaussian case.  }
\end{center}
\end{table}

\section{Discussion and Conclusions}
We considered two models for small primordial non-Gaussianity, one in which
the primordial density perturbation contains a term that is the square of a Gaussian
field (Model A), and one in which the primordial gravitational potential
perturbation  contains a term proportional to the square of a Gaussian (Model
B).
The non-Gaussianity of Model B is precisely that arising 
in standard slow-roll 
inflation and in non-standard inflation, while Model
A more closely resembles the non-Gaussianity that would be
expected from topological defects.
We investigated the relative sensitivities of several observables for  testing for deviations from
Gaussianity: CMB, LSS and  high-redshift and/or massive objects (e.g., galaxies and clusters). 

The analytic tools developed above allow us to address the question of whether
the abundance of currently known high-redshift objects can be accommodated
within the framework of inflationary models for a given cosmology.  Recently \scite{Willick00} has
studied in detail the mass determination of the cluster MS1054-03 concluding
that its mass lies in the range $1.4\pm 0.3 \times 10^{15}$ $M_{\odot}$ for
$\Omega_{\rm m}=0.3$ (similar to the independent mass estimates by,
e.g., \scite{Tranetal00} and \scite{Newmanarnaud00}). As already pointed out by \scite{Willick00}, for
$\Omega_{\rm m} \ge 0.3$ the expected number of objects like MS1054-03 in the
survey area is $ \le 0.01$; i.e., it must be a 3-$\sigma$ fluctuation or larger.
Using the formalism we have described here, a primordial non-Gaussianity parameterized by $\eps_{\rm B} \geq
400$ would be required to account for MS1054-03 as a  $1\sigma$
fluctuation in the $\Lambda$CDM model described above. This value is much too
large  to be consistent with slow-roll inflation. Our calculation shows that if
such a non-Gaussianity exists, it would be easily detectable from forthcoming CMB maps.

\section*{Acknowledgments}

LV and RJ thank the Caltech theoretical astrophysics group for hospitality. 
MK was supported in part by NSF AST-0096023, NASA NAG5-8506, and DoE
DE-FG03-92-ER40701.

\bibliographystyle{mnras}

\section*{Appendix}
In this Appendix we quote the expressions for the primordial  bispectrum and skewness for
the two non-Gaussian models considered in this paper.
The large-scale structure (LSS) bispectrum for model A is (e.g. VWHK00)
\be
B({\bf k}_1,{\bf k}_2,{\bf k}_3)=2\epsilon_{\rm A} P(k_1)P_(k_2)+cyc.
\ee
where $P$ denotes the power spectrum.
The cosmic microwave background (CMB) bispectrum for model A is (e.g. VWHK00)
\ba
B_{\ell_1\ell_2\ell_3}\simeq
\sqrt{\frac{(2\ell_1+1)(2\ell_2+1)(2\ell_3+1)}{4\pi}}
\left(^{\ell_1\ell_2\ell_3}_{0\;\; 0\;\; 0} \right)\nn
\times \frac{2\epsilon_{\rm
A}}{g}\left[\frac{2}{3}C_{\ell_1}C_{\ell_2}\frac{\ell_1^2\ell_2^2}{\ell_3^2}+cyc.\right]
\ea
where $C_{\ell}$ denotes the cosmic microwave background power spectrum,
$g$ denotes the radiation transfer function and $(\ldots)$ denotes the Wigner 3J
symbol.
The LSS bispectrum for model B is (e.g. VWHK00)
\be
B({\bf k}_1,{\bf k}_2,{\bf k}_3)\simeq \left[P(k_1)P(k_2) 2\epsilon_{\rm B}
\frac{{\cal M}_{k_3}}{{\cal M}_{k_1}{\cal M}_{k_2}}\right]+cyc.
\ee
where\footnote{This expression is strictly valid only for an Einstein de
Sitter Universe, for a more general model ${\cal M}$ is defined by
$\delta_{k}(z)={\cal M}_k(z)\Phi(k)$ where $\Phi$ denotes the gravitational
potential field. } ${\cal M}_{k}\sim(2k^2T(k)(1+z))/(3H_0^3)$ and $T$ denotes the
matter transfer function. 
The CMB bispectrum for model B is (e.g. \pcite{Luo94,WK00},VWHK00,\pcite{Komatsuspergel00}):
\ba
B_{\ell_1\ell_2\ell_3}=\sqrt{\frac{(2\ell_1+1)(2\ell_2+1)(2\ell_3+1)}{4\pi}}\left(^{\ell_1\ell_2\ell_3}_{0\;\;
0\;\; 0} \right)\nn
\times \frac{2\epsilon_{\rm B}}{g}(C_{\ell_1}C_{\ell_2}+cyc.)
\ea
 
The corresponding primordial skewness
$S_3=\langle\delta^3\rangle/\langle\delta^2\rangle^2$  where $\delta$ denotes
$\delta\rho/\rho$ for the large-scale structure case and $\Delta T/T$ for the
cosmic microwave background is easily obtained from the consideration that
$\langle\delta^3\rangle$ is given by:

\be
\langle\delta^3\rangle_{LSS}=\int \frac{d^3k_1}{(2\pi)^3}
\frac{d^3k_2}{(2\pi)^3}d^3k_3 B(\vk_1,\vk_2,\vk_3)\delta^D(\vk_1+\vk_2+\vk_3)
\ee
(in the absence of spatial filtering) and 

\ba
\langle\delta^3\rangle_{CMB}=\frac{1}{4 \pi}\sum_{\ell_1 \ell_2
\ell_3}\sqrt{\frac{(2\ell_1+1)(2\ell_2+1)(2\ell_3+1)}{4\pi}}\nn
\times \left(^{\ell_1\ell_2\ell_3}_{0\;\;
0\;\; 0} \right)B_{\ell_1\ell_2 \ell_3}
\ea

for LSS and CMB respectively.
For example in the large scale structure case, model A, for a power law power spectrum
and in the absence of spatial filtering\footnote{The expression for $\langle
\delta^3\rangle_{LSS}$ in  the general case can easily be derived following
the calculations of \scite{BK99} by setting $b_1=0$ and $b_2/2=\epsilon_{\rm A}$.} we have that $S_3=6\epsilon_{\rm A}.$ 
\end{document}